\documentclass[twocolumn,nofootinbib,prd,aps,superscriptaddress, amsmath,
tightenlines]{revtex4}
\usepackage{graphicx}

\begin{document}

\title{Exotic Implications of Electron and Photon Final States} 

\author{Lisa Randall}
\affiliation{Jefferson Laboratory of Physics, Harvard University
Cambridge, Massachusetts 02138, USA.}

\author{Mark B. Wise}
\affiliation{California Institute of Technology, Pasadena, CA 91125}

\date{July 9, 2008}

\begin{abstract}
New resonances with masses of order a few ${\rm TeV}$ might be discovered at the LHC. We show that no resonance that couples to electrons only through Standard Model interactions can decay to both $e^+e^-$and $\gamma \gamma$ with significant branching ratios.  This means that finding both electron-positron and two-photon final states is 
evidence that electrons couple directly to the new physics associated 
with the resonance and furthermore that the resonance is not spin-1.  The least fine-tuned such examples involve electron compositeness. One such example, Kaluza Klein excitations of the graviton in the version of the Randall Sundrum Model where Standard Model matter is located on the ${\rm TeV}$ brane, can be  distinguished from other possibilities by  its predicted branching fractions into the two modes. 
\end{abstract}

\maketitle

Direct experimental information on ${\rm TeV}$ scale physics will be possible for the first time at the ${\rm LHC}$. The hierarchy puzzle suggests that new physics beyond what is in the minimal Standard Model lurks at this scale. Of course the possible forms that new physics can take are constrained by precision electroweak data and flavor studies that have been performed at $B$ factories and elsewhere. It is nonetheless possible that new resonances with masses of order a ${\rm TeV}$ that are consistent with these constraints will be discovered at the ${\rm LHC}$. An interesting possibility that has been studied (for example, in \cite{CMS} and \cite{ATLAS}) is a very heavy resonance that decays to $e^+e^-$ and/or $\gamma \gamma$. The purpose of this  note is to show that powerful information on the nature of  such resonances can be learned from the study of $e^+e^-$ and $\gamma \gamma $ final states, particularly if they are found to have significant branching ratios to both of these modes.  In this note we demonstrate that if both modes are found with comparable branching ratios the electron must directly participate in the new physical interactions.

\begin{figure}
\centering
\includegraphics[trim= 0.1in 0.1in 0.1in 0.1in ]{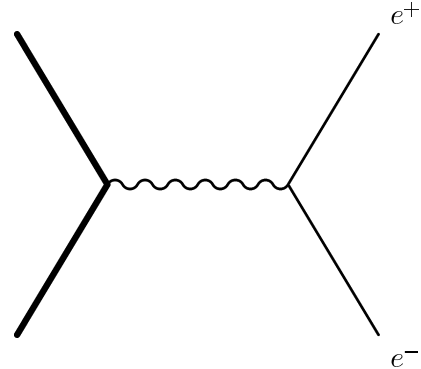}
\caption{Tree level diagram giving resonance decay to $e^+e^-$. New degrees of freedom, with $SU(2)\times U(1)$ quantum numbers, that are bound in the ${\rm TeV}$ mass scale resonance are denoted by the thick solid lines. The wavy line denotes a $\gamma$ or $Z$. The interactions that give rise to the bound-state are not shown.}\label{eedecay}
\end{figure}

\begin{figure}
\centering
\includegraphics[trim= 0.1in 0.1in 0.1in 0.1in ]{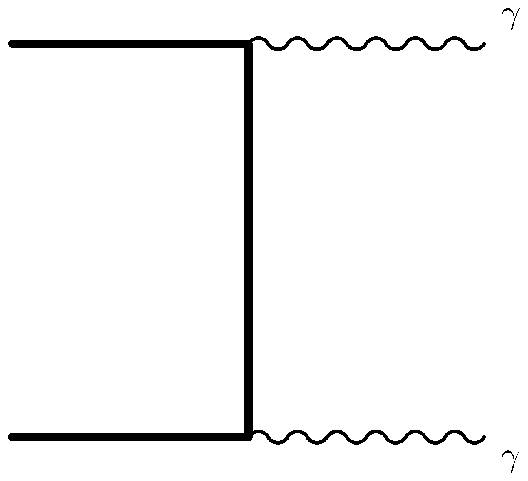}
\caption{{Tree level diagram giving resonance decay to $\gamma \gamma$. New degrees of freedom, with $SU(2)\times U(1)$ quantum numbers, that are bound in the ${\rm TeV}$ mass scale resonance are denoted by the thick solid lines. The wavy lines denotes a $\gamma$. The interactions that give rise to the bound-state are not shown.}}\label{ggdecay}
\end{figure}

New beyond-the-Standard-Model physics that is responsible for a ${\rm TeV}$ scale resonance does not necessarily  couple directly to electrons. If the resonance has constituents that carry $SU(2)\times U(1)$ quantum numbers it can decay to $e^+e^-$ and $\gamma \gamma$  at tree level  (see Fig.~(\ref{eedecay}) and Fig.~(\ref{ggdecay})) via Standard Model electroweak interactions and so significant branching ratios to either of these states is in principle possible. 

However, as we now show, if the electron mass is neglected it is never possible for such a resonance to decay at tree level via Standard Model electroweak interactions to both $e^+e^-$\footnote{Our conclusions apply to any charged lepton, in particular, $\mu^+ \mu^-$ but for simplicity we will refer only to electrons throughout.} and $\gamma \gamma$. This is true even if the interactions that bind the constituents of the resonance do not conserve $C$, $P$ or even $CP$.  Discovering both final states with comparable branching fractions will give valuable insights into the nature of the decaying resonance. In particular, if such decay modes are found,  the electron must couple directly to the source of new physics, a possibility that is severely constrained by limits on charged-lepton-flavor-changing processes. In fact, we argue that measuring the relative branching fractions can prove to be a powerful way of identifying the Randall Sundrum (RS) graviton even in the absence of spin measurements through angular distributions. 

We now demonstrate our result by considering sequentially decaying resonances of increasing spin, $J=0,1,2,...$ . (Fermionic resonances clearly cannot decay to $\gamma \gamma$ or $e^+e^-$.) The electrons in the Standard Model couple to the gauge bosons and the Higgs boson. However, the Higgs boson coupling is proportional to the electron mass and can be neglected since the Higss boson Yukawa coupling to the electron is of order $ 10^{-5}$.

A spin-zero resonance can decay to $\gamma \gamma$ at tree level via electroweak interactions when the resonance has charged constituents. For example, if there were no light quarks the charmonium resonance $\eta_c$ would have a large branching ratio to two photons. However a spin zero resonance decay amplitude to $e^+e^-$ is helicity-suppressed by a factor of the electron mass $m_e$.
For a spin zero resonance $R$ the tree level decay amplitude must take the form
\begin{equation}
\label{scalar}
{\cal A} \propto  \langle 0|J_{\mu} |R(p)\rangle D^{\mu \nu}(p){\bar u}(p)( a \gamma_{\nu} +b \gamma_{\nu}\gamma_5)u(p),
\end{equation}
where $D^{\mu \nu}(p)$ is a vector boson propagator, $J_{\mu}$ is the current it couples to, and we have allowed the vector boson to have both vector or axial vector couplings (or both) to the electrons. This formula applies with an intermediate $\gamma$ or $Z$ (or $Z'$ gauge boson if it exists). Since $\langle 0|J_{\mu} |R(p)\rangle \propto p_{\mu}$, $p_{\mu}D^{\mu \nu}(p) \propto p^{\nu}$ and $p^{\nu}{\bar u}(p)( a \gamma_{\nu} +b \gamma_{\nu}\gamma_5)u(p)=-2 m_e b{\bar u}(p)\gamma_5u(p)$ the amplitude ${\cal A}$ in Eq.~(\ref{scalar}) is proportional to the electron mass $m_e$. This supression of the decay amplitude persists at the loop level since in the massless limit gauge interactions preserve helicity. So although spin zero resonances can decay to photons, in the limit $m_e \rightarrow 0$, they cannot decay to $e^+e^-$ when the new physics, beyond that in the minimal Standard Model, does not couple directly to the electrons. 

There is a simple example of beyond-the-minimal-Standard-Model physics that can have a spin zero resonance with a significant branching ratio to $e^+e^-$, namely  a second  scalar doublet added to the Standard Model that does not have a vacuum expectation value and has an order unity coupling to the electrons. Such a model would in fact be quite unnatural, however, in that it would have to contain finely-tuned couplings so that in the fermion mass eigenstate basis it would not have unacceptably large flavor changing lepton couplings. Note that in this example the scalar has a small, loop suppressed, branching ratio for decay to two-photons.

For a spin-1 resonance, $R$, decay to $e^+ e^-$ at tree level is clearly possible. It could couple to a virtual $Z$ boson or photon. For example, the spin-one charmonium resonance $\psi(1S)$ decays with a $6\%$ branching ratio to $e^+e^-$ \cite{Yao:2006px}. However, Bose statistics (for the photons), rotational invariance and gauge invariance\footnote{We use the fact that the photon has only two transverse polarizations, which is a consequence of gauge invariance.} imply that a spin one resonance never decays to two photons~\cite{yang}.  A simple way to show this result (which is known as Yang's theorem) is to consider the possible forms of the decay amplitude.  The amplitude for $R(p) \rightarrow \gamma(k_1) \gamma(k_2)$ with $p=k_1+k_2$ must contain  each of the polarization vectors. We work in the rest frame of the decaying resonance where $p=(M,{\vec 0})$, $k_1=(M/2,{\vec k_1})$, $k_2=(M/2, {\vec k_2})$, $\epsilon(p)=(0, {\vec \epsilon})$, $\epsilon(k_1)=(0, {\vec \epsilon}_1)$ and $\epsilon(k_2)=(0, {\vec \epsilon}_2)$ and ${\vec k}_1=-{\vec k}_2$. Note that, ${\vec k}_{1,2}\cdot {\vec \epsilon}_{1,2}=0$ and $|{\vec k}_1|=|{\vec k}_2|=M/2$. Since all the dot products of momenta are fixed it suffices to consider terms with the least possible number of factors of momentum.
If the amplitude contains a Levi-Civita tensor then there are two cases to consider; either two or three of the polarization vectors must be contracted with its indices. First consider the case where two of the polarization vectors have indices contracted with the Levi-Civita tensor. Then Bose statistics plus rotational invariance implies that the amplitude has the form
\begin{equation}
\epsilon_{ijk}\epsilon_{1i}\epsilon_{2j}k_{1k} \vec{k}_1\cdot \vec{\epsilon}+\epsilon_{ijk}\epsilon_{2i}\epsilon_{1j}k_{2k} \vec{k}_2\cdot \vec{\epsilon}=0.
\end{equation}
If all the polarization vectors are contracted with indices in the Levi-Civita tensor the amplitude has the form
\begin{equation}
\epsilon_{ijk}\epsilon_{1i}\epsilon_{2j}\epsilon_k+\epsilon_{ijk}\epsilon_{2i}\epsilon_{1j}\epsilon_k =0.
\end{equation}
If the amplitude does not contain a Levi-Civita tensor then Bose statistics plus rotational invariance implies that it has the form,
\begin{equation}
{\vec \epsilon}_1 \cdot {\vec \epsilon}_2 {\vec k}_1 \cdot {\vec \epsilon}+{\vec \epsilon}_2 \cdot {\vec \epsilon}_1 {\vec k}_2 \cdot {\vec \epsilon}=0.
\end{equation}

So neither spin-0 nor spin-1 can decay to both electrons and photons without direct couplings of the new physics to electrons. It remains to consider spin-2 resonances and beyond. However, they cannot decay to $e^+ e^-$ through a virtual gauge boson because for such resonances $R$ the matrix element $\langle 0|J_{\mu} |R(p,\epsilon)\rangle$ vanishes since their polarization vectors have more than one Lorentz index and the momentum $p$ contracted with any of their polarization vector's indices gives zero. This result is consistent pattern  of a charmonium and bottomonium resonance decays;  no state with a comparable branching ratio to both $\gamma \gamma$ and $e^+e^-$ is observed.

Let's consider the implications of this result. The selection rule derived above does not forbid altogether a resonance $R$ from decaying to both $\gamma \gamma$ and $e^+ e^-$ with only Standard Model particles coupling directly to the electrons. Rather it forbids them from  decaying to the two final states at comparable rates. For example, suppose a spin-2 resonance has a tree level coupling to two photons. Then there is a one-loop diagram where the photons couple to electrons and produce the $e^+ e^-$ state. However since the $e^+ e^-$ state is produced at one loop while the decay to $\gamma \gamma$ proceeds at tree level the branching ratio to $e^+ e^-$ is very small.

\begin{figure}
\centering
\includegraphics[trim= 0.1in 0.1in 0.1in 0.1in ]{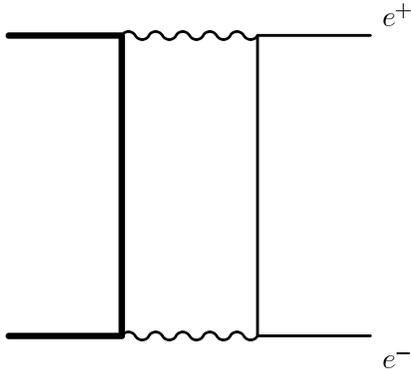}
\caption{{Possible one-loop diagram giving contributing to $R \rightarrow e^+e^-$. New degrees of freedom, with $SU(2)\times U(1)$ quantum numbers, that are bound in the ${\rm TeV}$ mass scale resonance are denoted by the thick solid lines. The wavy lines denote Standard Model $SU(2)\times U(1)$ gauge bosons. }}\label{loopdecay}
\end{figure}

Note that one cannot have measurable comparable branching fractions for electrons and photons occurring at the one-loop level either, since there will always be a dominant tree level decay to Standard Model particles as well. For a spin-1 resonance, the decay to two photons is forbidden by Yang's Theorem\footnote{The importance of this for resonances at the LHC was noted in \cite{CMS1}.}. For a spin-0 resonance, decay to $e^+e^-$ is helicity-suppressed. For the other possible spins, decay to $e^+ e^-$ does not occur at tree level (in the Standard Model interactions) through a virtual $\gamma$ or $Z$. However such resonances might decay to $e^+e^-$ at the one-loop level with intermediate Standard Model particles (e.g. W's Z's, or photons) that couple to the electron without a suppression from the electron mass (see Fig.~(\ref{loopdecay})).  However, with such interactions, the resonance can also decay at tree level to a final state with Standard Model gauge bosons. The branching ratio to  $e^+ e^-$, which proceeds at the loop level, will then be suppressed by $(g^2/{4 \pi} )^2\sim 10^{-5}$, where $g$ denotes either an $S(2)$ or $U(1)$ Standard Model gauge coupling constant.

We conclude that a significant branching ratio to both $\gamma \gamma$ and $e^+ e^-$ is possible only if new physics beyond that in the Standard Model couples directly to electrons.

Any new physics that couples to electrons is of course constrained by the absence of observed flavor-changing effects in the charged lepton sector ({ \it i.e.} the experimental limits on the branching ratios for, $\mu \rightarrow e \gamma$, $\tau \rightarrow e^+e^-e^-$, {\it etc.}). To naturally suppress flavor-changing effects of new physics that couples directly to  electrons requires a gauge principle.    Known examples would involve strong gauge interactions that couple both to the new resonance and to electrons.   Perhaps the best-motivated examples involve electron compositeness, with the new resonance and electrons bound by the same gauge interactions.  

In principle the new resonance could couple directly to electrons with a gauge principle protecting against flavor-changing effects. The only known such example would be a spin-2 particle with gravitational couplings. This occurs for KK resonances of the graviton. However, observable decays require TeV-scale suppressed interactions. The only known such example is the RS-model \cite{Randall:1999ee} where the Standard Model degrees of freedom are confined to the ${\rm TeV}$ brane.

In the effective four-dimensional theory the KK-graviton field $h^{\rm KK}_{\mu \nu}$ would couple to Standard Model degrees of freedom through the stress tensor~\cite{Davoudiasl:1999jd},
\begin{equation}
{\cal L}_{\rm int}= \kappa  h^{\rm KK}_{\mu \nu} T^{\mu \nu}_{\rm SM}
\end{equation}
where $T_{\rm SM}$ is the Standard Model energy-momentum stress tensor and $\kappa$ is a coupling constant that depends on parameters in the RS model. From
the dual perspective this is an example of the composite electrons suggested above. 

In this case, there is
  a significant branching ratio to both $e^+ e^-$ and $\gamma \gamma$.  
 Neglecting the final state masses the rate for the KK-graviton to decay to particular Standard Model final states is
\begin{equation}
\label{rate}
\Gamma={\kappa^2f M^3\over 80 \pi},
\end{equation}
where $M$ is the mass of the KK-graviton and $f$ is a factor that depends on the final state. The values of $f$ for all the possible Standard Model two body final states  are given in Table 1. We note that the branching ratios into the various final states are the same as those that  would apply for a graviton in the flat-extra-dimensional scenario so this table has no numbers that cannot be obtained from that Ref.~\cite{han}. Our result is the selection rule that states that such comparable branching ratios will not occur in any scenario where the electron does not couple directly to new physics. Furthermore, RS is the only scenario where we expect the KK graviton to decay in the detector so that these branching fractions could be measurable.

Although any model of composite electrons might allow comparable $e^+ e^-$ and $\gamma \gamma$ final states, the KK-graviton should be readily identifiable when both branching fractions are measured, since the ratio of branching ratios to these final states is predicted to be ${\rm Br}(h^{\rm KK} \rightarrow \gamma \gamma)/{\rm Br}(h^{\rm KK} \rightarrow e^+ e^-)=2$ \cite{han}. This is very important in that spin measurements through angular dependence of the final state particles can be quite challenging and will certainly require more statistics than discovery and measuring relative branching fractions. Randall-Sundrum KK-gravitons have been searched for in these channels at the Tevatron \cite{Abazov:2005pi}.
{
\begin{center}
{\begin{tabular}{lcccccccc}
\hline
&$gg$& $\gamma \gamma$ & $ZZ$& $WW$ &$\bar q q$ &${\bar l} \l$ & $hh$ & ${\bar \nu} \nu$ \\
\hline
${f}$: & ${8}$ & ${1}$ & ${1}$ & ${2}$ &${3/2}$ &${1 /2}$& ${1/ 6}$ & ${1 / 4}$  \\
\hline 
\end{tabular}}
\end{center}}
\noindent Table~1: Factors $f$ for various two-body standard model final states.
\vskip0.15in
\noindent In the Table 1 $q$ denotes a single quark species so to get the total rate to quark-antiquark pairs one must muliply Eq.~(\ref{rate}) by a factor of $6$, $l$ denotes a single charged lepton species so to get the total rate to charged lepton-antilepton pairs one must multiply Eq.~(\ref{rate}) by a factor of $3$ and $\nu$ denotes a single left handed neutrino species (we assume any right handed neutrinos are too heavy to be produced in the decay). If only Standard Model final states are possible then the above results imply that $ {\rm Br}( h^{\rm KK}\rightarrow e^+e^-) \simeq 0.022$ and ${\rm  Br}( h^{\rm KK}\rightarrow \gamma \gamma) \simeq 0.044$.

In conclusion, measuring branching the fractions for resonant decay to $e^+e^-$ and $\gamma \gamma$ can tell us that the electron couples directly to new, beyond-the-Standard Model physics, providing indirect evidence that the electron is composite, and furthermore serve as a way of identifying the RS graviton KK mode for brane-localized fermions and gauge bosons. 

\begin{acknowledgments}
We thank Harvey Newman and Marat Gataullin for alerting us to the photon measurements and for proofreading the manuscript.
This work was supported in part by DOE grant DE-FG03-92ER4070 and  by NSF under grants
 PHY-0201124 and PHY-055611. This work was completed while LR was a Moore
Distinguished Scholar at Caltech. 
\end{acknowledgments}


\end{document}